\documentclass[aps,showpacs,amsmath,amssymbd,dvips,showkeys,preprint,preprintnumbers]{revtex4}
\usepackage{amsfonts}

\usepackage{subfigure}
\usepackage{amsmath}
\usepackage{graphicx}
\usepackage{color}

\def \be  {\begin{equation}}
\def \ee  {\end{equation}}
\def \ee  {\end{equation}}
\def \bea {\begin{eqnarray}}
\def \eea {\end{eqnarray}}

\def \Tr  {\bf{Tr}}

\begin{document}

\preprint{ECTP-2015-06}
\preprint{WLCAPP-2015-06}
\vspace*{.3cm}

\title{Lattice QCD thermodynamics and RHIC-BES particle production within generic nonextensive statistics}

\author{Abdel Nasser  TAWFIK}
\email{a.tawfik@eng.mti.edu.eg}
\affiliation{Egyptian Center for Theoretical Physics (ECTP), Modern University for Technology and Information (MTI), 11571 Cairo, Egypt}
\affiliation{World Laboratory for Cosmology And Particle Physics (WLCAPP), Cairo, Egypt}

\begin{abstract}

The current status of implementing Tsallis (nonextensive) statistics on high-energy physics is briefly reviewed. The remarkably low freezeout-temperature, which apparently fails to reproduce the first-principle lattice QCD thermodynamics and the measured particle ratios, etc. is discussed. The present work suggests a novel interpretation for the so-called {\it "Tsallis-temperature"}. It is proposed that the low Tsallis-temperature is due to incomplete implementation of Tsallis algebra though exponential and logarithmic functions to the high-energy particle-production. Substituting Tsallis algebra into grand-canonical partition-function of the hadron resonance gas model seems not assuring full incorporation of nonextensivity or correlations in that model. The statistics describing the phase-space volume, the number of states and the possible changes in the elementary cells should be rather modified due to interacting correlated subsystems, of which the phase-space is consisting. Alternatively, two asymptotic properties, each is associated with a scaling function, are utilized to classify a generalized entropy for such a system with large ensemble (produced particles) and strong correlations. Both scaling exponents define equivalence classes for all interacting and noninteracting systems and unambiguously characterize any statistical system in its thermodynamic limit. We conclude that the nature of lattice QCD simulations is apparently extensive and accordingly the Boltzmann-Gibbs statistics is fully fulfilled. Furthermore, we found that the ratios of various particle yields at extreme high and extreme low energies of RHIC-BES is likely nonextensive but not necessarily of Tsallis type.

\end{abstract}


\maketitle
\tableofcontents

\section{Introduction}
\label{sec:introduction}

The extensivity of the entropy of a system consisting of two initially isolated, sufficiently separated subsystems ($A$ and $B$), with $\Omega_A$ and $\Omega_B$ being the respective numbers of states, is defined  as  $S(\Omega_{A+B})=S(\Omega_A)+S(\Omega_B)$, while the additivity as $S(\Omega_A\, \Omega_B)=S(\Omega_A)+S(\Omega_B)$ \cite{Thurner1,Thurner2}. Only if $\Omega_{A+B}=\Omega_A\, \Omega_B$, both extensivity and additivity coincide  \cite{Thurner1}. Boltzmann-Gibbs entropy, $S[p] = \sum_{i}^{\Omega} g(p_i)$, where $g(p_i)=−p_i\, \ln p_i$, fulfils both  property. 

In understanding many features of high-energy collisions and lattice QCD simulations, the hadron resonance gas (HRG) model is widely implemented as a successful approach possessing - among others - the correct degrees of freedom \cite{Karsch:2003vd,Karsch:2003zq,Redlich:2004gp,Tawfik:2004sw,Tawfik:2004vv}. Also, this type of thermal-statistical approaches play an essential role in characterizing the extensive statistics in high-energy collisions. Thus, we focus the discussion on HRG, in which the constituents (the hadrons and resonances) are taken as point-like. This is an ideal (collision-free) gas. In Boltzmann-Gibbs (BG) statistics, the HRG grand-canonical partition-function is given by Hamiltonian and the baryon number operators,  $\mathbf{\hat{H}}$ and $\mathbf{\hat{b}}$, respectively, and depends on the temperature ($T=1/\beta$) and the chemical potentials ($\mu$), 
\bea \label{eq:zTr}
Z(\beta,V,\mu) &=& \Tr \, \left[\exp^{\beta (\mu \hat{b}-\hat{H})}\right].
\eea
This can be characterized by various but a complete set of microscopic states and therefore the physical properties, including entropy, can be estimated from additive contributions from uncorrelated {\it independent} constituents (the hadrons and resonances). The main motivation of using the Hamiltonian is that it contains all relevant degrees of freedom of confined, strongly interacting hadron matter, i.e., the HRG applicability is limited below the critical temperature. Furthermore, the Hamiltonian implicitly includes the interactions that result in the formation of new resonances \cite{hagedorna,hagedornb}. These are nothing but the strong interactions \cite{Tawfik:2004sw}. 

On the other hand, the hadron resonances treated as a free gas \cite{Karsch:2003vd,Karsch:2003zq,Redlich:2004gp,Tawfik:2004sw,Tawfik:2004vv} are assumed to add to the thermodynamics of the hadron matter. For an ideal gas composed of hadron resonances with masses $\le 2~$GeV~\cite{Tawfik:2004sw,Vunog}, the correlation can be taken into consideration and the resulting partition function gives a quite satisfactory description for the particle production in heavy-ion collisions and the lattice QCD thermodynamics \cite{Karsch:2003vd,Karsch:2003zq,Redlich:2004gp,Taw3b,Taw3c}. This mass limit is set in order to avoid Hagedron singularity \cite{Tawfik:2004sw}. 

Therefore, HRG thermodynamics including entropy can be deduced from the sum over stable hadrons and resonances
\bea
\ln Z(T,V,\mu) &=& \sum_i\pm \frac{g_i}{2\pi^2}\,V\int_0^{\infty} k^2 dk \ln\left(1\pm \, e^{(\mu-\epsilon_i(k))/T}\right), \label{eq:PFq1}
\eea
where $\pm$ stands fermions and bosons, respectively.  In other words, the HRG model implements an additive approach. The resulting entropy is resulted from a sum over entropies of each constituent. This implies that the uncorrelated constituents of the HRG approach assure additivity, as well. Extensivity coincides with additivity if the number of correlated and that of uncorrelated states are identical, which seems to be the case in an ideal HRG. But in an interacting statistical system the phase-space is only partly visited \cite{Tawfik:2010kz,Tawfik:2010pt} and the number of uncorrelated (noninteracting) is less than that of correlated states \cite{Thurner2}, so that the additive entropy becomes no longer extensive and vice versa. This highlights the need for a proper implementation of nonextensivity in the  HRG model. 

As shall be discussed, applying Tsallis algebra, i.e., expressing exponential and logarithmic functions in Tsallis algebra \cite{Tsallis1988,PratoTsallis1999,Tsallis2005}, to the HRG partition function generates the well-known $q$-terms. This one-to-one substitution apparently doesn't assure a full incorporation of nonextensivity. Firstly, it makes the new version of HRG no longer able to reproduce the first-principle lattice QCD thermodynamics. Secondly, the resulting freezeout temperature is very low. Even when the transverse momentum spectra, which are considered as a best implication of Tsallis statistics on high-energy collisions \cite{worku1,Cleymans2014}, are fitted, the freezeout temperature is remarkably low. These problems rely on the improper implementation of Tsallis statistics to the high-energy particle production rather than to short-cuts in this type of statistics, which is - no doubt - an essential generalization to the Boltzmann-Gibbs (BG) statistics. The present work is not criticizing Tsallis statistics, itself. In contrary it highlights the importance of the proper implementation. In emphasizing this,  a short but comprehensive review is firstly introduced. Furthermore, a new approach, generic nonextensive statistics, is proposed. Accordingly, no longer low temperature is obtained but remarkable well both lattice QCD thermodynamics and the RHIC-BES particle ratios are reproduced.

A nonextensive generalization of Hagedorn's theory was proposed long time ago \cite{beck1}. Associating local temperature fluctuations with $q$-exponentials observed in high-energy scattering data \cite{beck2} and the suggestion of superstatistics rather than Tsallis statistics that could be relevant in high-energy physics have been discussed recently \cite{beck3}.

It is noteworthy highlighting that the nonextensive thermodynamics at finite baryon chemical potential, known as chemical freeze-out diagram, is not at all compatible with the one drawn from the thermal-statistical models \cite{Tawfik:2014eba} and with the available lattice QCD calculations \cite{Depman2015}. The resulting freezeout temperatures are much lower than whose deduced from thermal-statistical and lattice QCD approaches. The great attempts to extrapolate the resulting freezeout temperature (when implementing Tsallis statistics) to the well-know BG-temperature shall be discussed, shortly. The latter is compatible with the first-principle lattice QCD calculations, for instance. We introduce generic nonextensive statistics, in which both Tsallis (nonextensive) and Boltzmann-Gibbs (extensive) are very special cases. In this generic approach, the degree of (non)extensivity is effectively dynamically determined by the phase-space distribution-function. The latter represents the main aspect reflecting the consequences of nonextensivity in high-energy particle production rather than just $q$-algebra, for instance.

The present work is organized as follows. A brief reminder to the so-far implications of nonextensive statistics (Tsallis) on high-energy collisions is given in section \ref{sec:nonexrensive}. A very short review on Tsallis temperature, which is deduced when confronting thermal models with Tsallis statistics to the particle production and transverse momentum spectra, shall be outlined in section \ref{sec:TsllsT}. The Tsallis algebra (exponential and logarithmic functions) and the resulting distribution functions shall be elaborated in section \ref{sec:TsallisAlgebra}. As an alternative or better to say a generic approach, we propose to utilize the possible connection between nonextensivity and the possible modification in the phase-space in section \ref{sec:phasespace}. The generic axiomatic entropy derivation and the distribution functions, which take into consideration such possible modifications are reviewed in section \ref{sec:axiomatic}. As examples on this approach, we discuss both Ising model, section \ref{sec:ising} and coalescence quark-model, section \ref{sec:coalescent}. Section \ref{sec:pfunction} is devoted to the generic nonextensive partition function and its results. The results and conclusions shall be given in section \ref{sec:results} and section \ref{sec:conclusions}, respectively.

\section{Reminder to nonextensive statistics in high-energy collisions}
\label{sec:nonexrensive}

Various approaches of nonextensive statistics have been implemented on high-energy collisions.
\begin{itemize}
\item{Central limit theorem (CLT) in $q$-algebra}
 as statistics and probability theory plays an essential role in various applied sciences, including statistical mechanics. CLT is well established for weakly dependent random variables \cite{Doukhan1994,DehlingDenkerPhilipp1986} and thus might not be applicable in high-energy physics, where essential variables such as chemical freezeout and temperature are strongly correlated with each other and with other variables such as volume and beam energy. Gaussian distributions as an attractor of independent systems with a finite second variance were first shown by de Moivre, de Laplace, Poisson and Gauss and later on by Chebyshev, Markov, Liapounov, Feller, Lindeberg, and L\'evy considerably to contribute to the development CLT. In classical BG statistical mechanics, it is well known that Gaussian distributions maximize the entropy $S_{BG}=-\sum_i p_i \ln p_i$ or $-\int  dx \, p(x) \ln p(x)$ in the continuous form. Obviously, these are rather extensive.

\item{$q$-generalization (or CLT-analogue):} 
CLT was the basis for the generalization of BG statistics \cite{Tsallis1988}. In its continuous limit,
\bea
S_q &=& \frac{1-\sum_i\, p_i^q}{q-1} = \frac{1-\int  dx \, [p(x)]^q}{q-1}, \label{eq:Sq1}
\eea 
where $q\in \mathcal{R}$. At $q=1$, the entropy reaches its maximum in $q$-Gaussian distribution, $S_1=S_{BG}$ \cite{PratoTsallis1999}.  Besides these assumptions, conjectures \cite{Tsallis2005}, numerical indications \cite{MoyanoTsallisGellmann2006}, and recent developments \cite{Tsallis1988,PratoTsallis1999,TsallisBukman}, $q$-analogue for CLT, such as $q$-sum and $q$-product was proposed. This is nothing but Tsallis statistics. The $q$-algebra, the Tsallis-algebra, relies on $q$-exponential and $q$-logarithm, which are based on conversion into power-law scales:
\bea
\exp_q(x) &=& [1+(1-q)x]^{1/(1-q)}, \label{eq:exp}\\
\ln_q(x) &=& \frac{x^{1-q}-1}{1-q}, \label{eq:ln}
\eea
where $x>0$. With applying Tsallis-algebra in HRG partition function, we mean substituting its $\exp$ and $\ln$ terms by Eq. (\ref{eq:exp}) and Eq. (\ref{eq:ln}), respectively.
\end{itemize}

As a short review on so-far implications of nonextensive statistics on high-energy physics, we mention three examples, in which $q$-statistics (Tsallis statistics) is utilized \cite{Tsallis1988,PratoTsallis1999,TsallisBukman}.
\begin{itemize}
\item{Transverse momentum distribution ($p_T$):} It was concluded that a thermodynamically consistent $q$-version of $p_T$-distribution describes well the transverse momentum spectra of the well-identified particles with fitted temperature ($T_q$) and baryon chemical potential ($\mu_B$)  \cite{worku1}
\begin{equation}
\lim_{q\rightarrow 1}\frac{d^{2}N}{dp_T~dy} = 
\frac{g\, V}{(2\, \pi)^2}\, p_T\, m_T\, \cosh y\;
\exp\left(-\frac{m_T\, \cosh y -\mu_B}{T_q}\right), \label{boltzmann}
\end{equation}
where the subscript $q$ refers to $q$-statistics. 
At LHC, typical values for $q$ and $T_q$ range from $1.1$ to $1.17$,  and $70$ to $90~$MeV, respectively \cite{Cleymans2014}.  Almost same values are obtained in Ref. \cite{deppman2012,sena}. At LHC energy, the resulting baryon chemical potential is negligibly small. Accordingly, such nonextensive approach with even thermodynamical self-consistency predicts a limiting effective temperature and a limiting entropic index and apparently imposes a limited applicability of Tsallis statistics in high-energy collisions \cite{DepmannHagedorn}. A little bit higher $q$ and $T_q$ values have been obtained for $p_T$ distributions of hadrons produced from $e^+e^-$ collisions \cite{Bediaga}.

\item{Hagedorn-Tsallis distribution:}  
It was shown that the HRG model, in which  two exponentially-increasing mass spectrum functions are included [not from the particle data group (PDG)], describes well the lattice thermodynamics \cite{deppman2012,DepmannHagedorn,latticeQCD}. Even here, the resulting temperature and entropic index are very low, $60.7\pm0.5~$MeV, and $1.138\pm0.006$, respectively. 

\item{Nonextensivity at finite chemical potential:} 
The grand-canonical partition function for nonextensive ideal quantum gas is given as \cite{Depman2015} 
\begin{eqnarray}
\log \Xi_q (V, T, \mu) = \mp\, V\, \int \dfrac{d^3p}{(2 \pi)^3} \sum_{r=\pm} \Theta(rx)\, \log_{q}^{(-r)} \left( \dfrac{ e_{q}^{(r)}(x) \mp 1}{ e_{q}^{(r)}(x)} \right), \label{eq:nonextnsDeppmann}
\end{eqnarray}
where top sign stands for bosons and bottom one for fermions, $x=\beta (E(p) - \mu_B)$, with $E(p)=\sqrt{p^2 + m^2}$, $m$ being the hadron mass, and $\Theta$ is a step function assuring real dispersion relation. For bosons, this expression is only defined at $\mu_B\leq m$. Therefore, the term with negative $r$ is applied for fermions, explicitly, but it contributes at $\mu_B>m$, as well. The results on transverse momentum distribution, average particle number, energy density and entropy are almost identical to the ones obtained in Refs. \cite{worku1,DepmannHagedorn}, when $x \geq 0$. However, some differences have been obtained at $x<0$ \cite{Depman2015} and some severe results on $q$-statistics are depicted in Fig. \ref{fig:fg1}. On the other hand, it should be highlighted that as $q\rightarrow 1$, Eq. (\ref{eq:nonextnsDeppmann}) reduces to the well-known Fermi-Dirac and Bose-Einstein partition functions.

\end{itemize}

\subsection{A short review on extrapolation Tsallis to BG freezeout temperature}  
\label{sec:TsllsT}
  
As discussed in the previous sections, the resulting low temperature is due to incomplete implementation of Tsallis statistics on high-energy physics. The proper implementation should not explicitly focus on substituting the exponential and logarithmic terms in the distribution function by Eq. (\ref{eq:exp}) and Eq. (\ref{eq:ln}), respectively, as the case so far. The possible modifications in the phase space itself should also be taken into consideration. 

Despite the absence of a solid physical interpretation for the resulting freezeout temperature, except an inexplicable one based on {\it ab initio} assumed difference between BG- and Tsallis-statistics, an extrapolation to the BG freezeout temperature ($T_{ch}$), which is deduced from the fit of various particle ratios with the {\it extensive} thermal-statistical models, for instance, reads \cite{deppman2012}
\bea
T_q &=& T_{ch} + (q-1)\, k,
\eea
where the constant $k$ depends on the so-called energy transfer between source and surrounding.  At $T_{ch}=192\pm 15~$MeV, $k=-(950\pm10)~$MeV \cite{sena}. For a recent review, the readers are kindly advised to consult Ref. \cite{Wilk2016}.

Another interpretation for the reported difference between freezeout temperature at $q\neq1$ and that at $q=1$ was proposed, recently \cite{Biro}.  From a physical model of a finite thermostat, 
\bea
T_{ch} &=& T_q\, \exp\left(-\frac{S_{q}}{C}\right),
\eea
where $S_q$ is the nonextensive entropy and $C$ is the heat capacity of reservoir system. 

As mentioned in earlier sections, the $q$-temperature is $\sim 2-3$ times smaller than the BG freezeout value ($T_{ch}$). Accordingly, a remarkable difference was found between {\it nonextensive} freezeout diagram and the one determined from the {\it extensive} statistical fit of various measured particles ratios \cite{worku1,DepmannHagedorn,Depman2015}, Fig. \ref{fig:fg1}. 

Also at Tsallis temperature, the resulting QCD thermodynamics  considerably differs from the first-principle lattice calculations, which  - in tern - well reproducible by the HRG model,  Fig. \ref{fig:fg2a}, for instance.
  
In the present work, we introduce a generic (non)extensive approach taking into consideration the possible modification in the phase space and measuring the degree of (non)extensivity, simultaneously. Accordingly, the resulting freezeout temperature in found compatible with both the one deduced from the lattice calculations and the statistical fits of various particle ratios. Furthermore, the degree of (non)extensivity is practically determined by two critical exponents, which define equivalence classes for all interacting and noninteracting systems and unambiguously characterize any statistical system in its thermodynamic limit.

\subsection{Nonextensivity and nonexponential canonical distributions}
\label{sec:nonexp}

Here, we give a short list of nonextensive entropies leading to nonexponential canonical distributions \cite{Thurner2,Biro2015}. 
\begin{itemize}
\item Renyi entropy is considered as a generalisation of Boltzmann-Gibbs-Shannon entropy \cite{renyi1961}:
\begin{eqnarray}
S_R &=& \frac{1}{1-q} \ln \sum_i^{\Omega} p_i^q,
\end{eqnarray}
where $p_i$ is the probability of $i$-th state. Here, the parameter $q$ takes arbitrary values $0<q<\infty$, including $q=1$ which is supposed to characterize the BG entropy. While Boltzmann-Gibbs-Shannon entropy is strongly related to thermodynamics, only extreme cases of Renyi entropy ($q\rightarrow 0$ and $q\rightarrow \infty$, known as min- and max-entropy, respectively) find a convincing use in thermodynamics. Thus, we conclude that $S_R$ increases with increasing $q$.
\item Tsallis nonextensive entropy \cite{Tsallis1988}:
\begin{eqnarray}
S_T &=& \frac{1}{1-q} \ln \sum_i^{\Omega} \left(p_i^q - p_i\right).
\end{eqnarray}
Details about Tsallis entropy and its implications on high-energy collisions can be taken from Refs. \cite{Tsallis1988,PratoTsallis1999,Tsallis2005,worku1,Cleymans2014}.
\item Generalized entropies assuring (non)extensivity that might arise - among others - from interaction and correlation in the system of interest, section \ref{sec:axiomatic},  \cite{Thurner2}
\begin{eqnarray}
S_q[p] &=& \sum_{i=1}^{\Omega} g(p_i).
\end{eqnarray}  
Here, the function $g$ can have any functional form, for instance incomplete gamma function. The degree of (non)extensivity is restrictively determined by the four Khinchin axioms \cite{ShannonKhinchina,ShannonKhinchinb}, which are given as follows.
\begin{enumerate}
\item Continuity: for any $n\in  \mathbb{N}$, $S_1(p)$ becomes continuous with respect to $p\in\Delta_n$. This means that the entropy contentiously depends on $p$ and consequently the function $g(p)$ is a continuous function.
\item Maximality: for given $n\in  \mathbb{N}$ and for $(p_1,\cdots,p_n)\in\Delta_n$, $S_1(p_1,\cdots,p_n)$ becomes maximum at $p_i=1/n$, where $i=1,\cdots,n$. This means that the entropy is maximum for equi-distribution ($p$) and consequently the function $g(p)$ is a concave function.
\item Expandability: $S_1(p_{1},\cdots,p_{n},0) = S_1(p_{1},\cdots,p_{n})$. This means that adding a non-zero probability state does not change the entropy, $g(0)=0$, i.e., the entropy is invariant under adding zero-probability states to the system.
\item Generalized Shannon additivity: if $p_{ij} \geq 0$, $p_i=\sum_{j=1}^{m_i} p_{ij}$ and $\sum_{i=1}^{n} p_{i}=1$, where $i=1,\cdots,n$ and $j=1,\cdots,m_i$, then $S(p_{11},\cdots,p_{nm_n}) = S_1(p_{1},\cdots,p_{n}) + \sum_{i=1}^{n} p_i S_1\left(\frac{p_{i1}}{p_i}\cdots,\frac{p_{i m_i}}{p_i}\right)$. This means that $S_n$ is the Shannon entropy given by the class (2), i.e., the entropy of a system which is divided into sub-systems $A$ and $B$ is given as $S_A$ plus the expectation value of $S_B$ conditional on $A$.
\end{enumerate}
The extensivity is conditioned upon fulfilling the four axioms, while the nonextensivity upon violating the fourth axiom, separability. 
\end{itemize}

\section{Nonextensive statistics in high-energy physics}

\subsection{Tsallis entropy and distribution functions}  
\label{sec:TsallisAlgebra}

\begin{figure}[!htb]
\centering{
\includegraphics[width=8.cm]{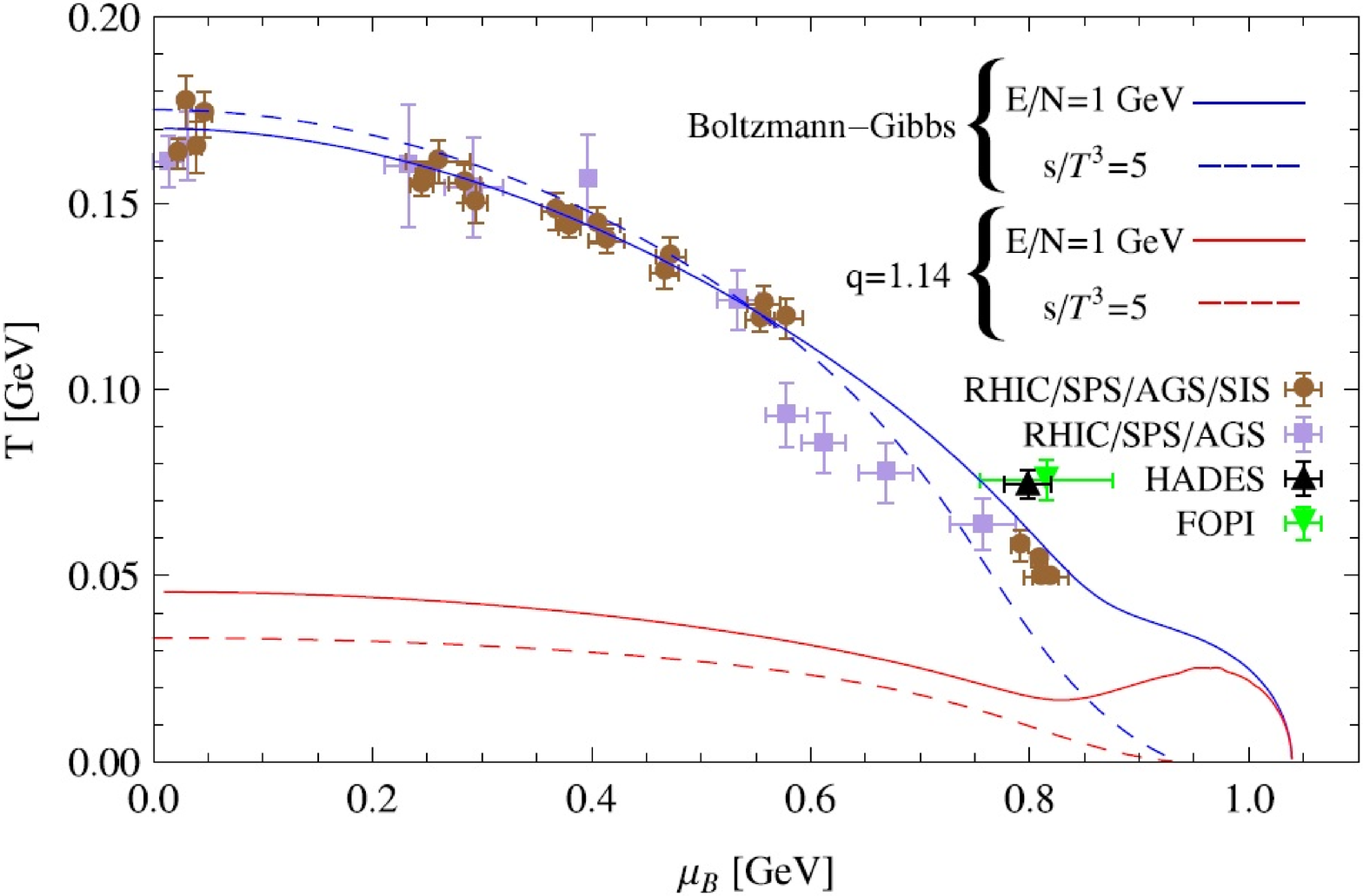}
\includegraphics[width=8.cm]{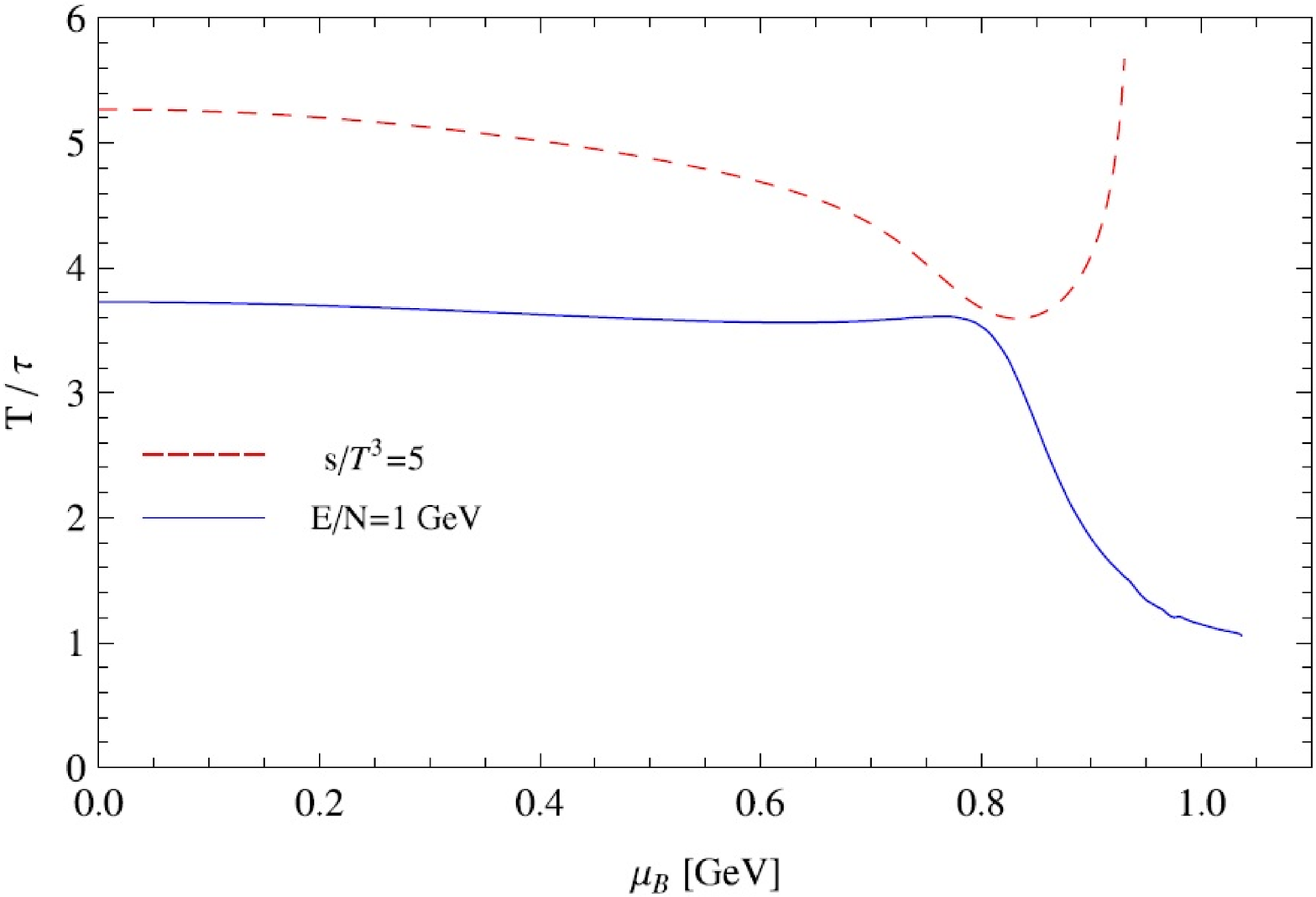}
\caption{Left-hand panel presents the freezeout diagram as temperature $T_{ch}$ versus baryon chemical potential ($\mu_B$). Symbols stand for phenomenologically-deduced parameters from the confrontation of BG thermal models and various particle ratios measured in different experiments. Top solid and dashed curves depict the results from BG thermal models at $E/N=1~$GeV and $s/T^3=5$, respectively. The results from nonextensive statistics, for instance, Eq. (\ref{eq:nonextnsDeppmann}) are given by the bottom curves. The freezeout temperature (here called $T$) relative to the $q$-value of temperature (called $\tau$, here) resulted from $E/N=1~$GeV (solid curve) and $s/T^3=5$ (dashed curve) are presented in right-hand panel. The graphs are taken from \cite{Depman2015}.
\label{fig:fg1}
}}
\end{figure}

Through substituting $\exp$- and $\ln$-functions by their counterparts in Eq. (\ref{eq:exp}) and Eq. (\ref{eq:ln}), respectively,  it is said that $q$-statistics is implemented in the HRG model, for instance. Fig. \ref{fig:fg1} presents the freezeout diagram, $T_{ch}$ vs. $\mu_B$, calculated from the HRG model, in which $q$-statistics is utilized \cite{Depman2015}. Symbols stand for the phenomenologically-deduced freezeout parameters from the confrontation of BG thermal models ({\it extensive}) and  various particle ratios measured in different experiments \cite{Tawfik:2014eba}. Top solid and dashed curves give the results from BG thermal models at two freezeout conditions; $E/N=1~$GeV (solid curve) and $s/T^3=5$ (dashed curve), respectively. The results from nonextensive statistics, for instance, Eq. (\ref{eq:nonextnsDeppmann}), are given by the bottom curves. The freezeout temperature (here called $T$) relative to the $q$-temperature (here called $\tau$) resulted from the condition $E/N=1~$GeV (solid curve) and from the condition $s/T^3=5$ (dashed curve)  are presented in right-hand panel. From $E/N=1.0~$GeV, it is obvious that at $\mu_B\lesssim 0.8~$GeV, $T_{ch}$ is $\sim 4$ times higher than $T_q$. At higher $\mu_B$, $T_{ch}\rightarrow T_q$. From the condition $s/T^3=5$, at $\mu_B\lesssim 0.5~$GeV, $T_{ch}/T_q\simeq 5$. At $\mu_B\sim 0.8~$GeV, $T_{ch}\rightarrow\, \sim 4\, T_q$. At higher $\mu_B$, $T_q\rightarrow \infty$. These results illustrate the great difference in the  $T$ - $\mu$ freezeout diagram from nonextensive and extensive approaches.

\subsection{Nonextensivity from phase-space modification}  
\label{sec:phasespace}

In high-energy collisions and lattice QCD thermodynamics, the great discrepancy between their results at $q=1$ and at $q\neq1$  might be interpreted due to differences between power laws imposed by the $q$-statistics and the Boltzmann factor in $q=1$ quantum and classical statistics. Furthermore, even the remarkable success of  the $q$-statistics in describing the transverse momentum spectra, apart from the noticeably-low temperatures, imposes an {\it ad hoc} factorization from Tsallis statistics, which is only valid for many microstates, down to one-particle problem. Not only this but accordingly, a mean value for $q$ is implemented instead of the standard average values.  In other words, $q$-statistics should be applied to systems with many microstates. Accordingly, the parameter $q$ might be varying. One should assume mean values of $q$ instead of the average ones. Again, the proper implementation of Tsallis statistics should be restricted to micro or even single states and $q_i$ should be first assigned to each $i$-th micro or single state. It is worthwhile emphasizing that even the mathematically proper factorizations \cite{factrza,factrzb,factrzc} are nothing but approximative approaches to the exact Tsallis statistics.

The strongly interacting (correlated) systems of high-energy collisions remain strongly correlated, even at temperatures greater than the critical one. Such QCD systems are characterized by long-range and strong interactions and can change their properties, symmetries and degrees of freedom \cite{Thurner2}. These are statistical systems violating one of the four Shannon-Khinchin axioms \cite{ShannonKhinchina,ShannonKhinchinb}. In order to treat such systems, thermodynamically, entropic nonextensivity should be assured. $q$-entropy, Eq. (\ref{eq:Sq1}), is just one possibility. These are examples on argumentations why generic nonextensive statistics should be applied in, section \ref{sec:axiomatic}.

As mentioned in earlier sections, nonextensivity is only applicable whenever microscopic states are correlated. Absence of correlations (interactions) means that the states are independent on each others. Replacing exponential and/or logarithmic terms of the distribution function by $q$-expressions as done so far when Tsallis algebra, Eq. (\ref{eq:exp}) and Eq. (\ref{eq:ln}),  has been applied so far to the high-energy particle-production \cite{worku1,DepmannHagedorn,Depman2015}, seems not fully characterizing the correlations or the interactions in such strongly interaction systems, which can partly be described by HRG but fully characterized by QCD. Such strongly correlated systems are likely associated with remarkable changes in the phase space, itself \cite{Tawfik:2010kz,Tawfik:2010pt}. These are additional argumentations why the generic nonextensive statistics, section \ref{sec:axiomatic}, is proper to be utilized. 
  
This motivated our discussion on how the statistics describing the phase-space volume, the number of states and the elementary-cell changes when switching on the interactions (no longer Boltzmann subsystems) become correlated. If equal and distinguishable $N$-subsystems, of which the phase space is consisting, are locally correlated, {\it extensive} BG-entropy can be applied \cite{GellMannTsallis2005}. If such subsystems are even independent, the BG entropy becomes additive, as well. On the other hand, if they are globally correlated, a large class of systems can be expected; $q\neq 1$ (Tsallis) represents a special case. But an asymptotic scale-freedom is likely when all marginal probabilities of discrete binary-subsystems have been reached. It has been shown that such duality fixes a unique escort probability \cite{ThurnerGellMann2012}. The complexity of generalized entropy dependence on the phase space is not given by just the binary correlations.

\subsection{Axiomatic generic entropy and distribution functions}
\label{sec:axiomatic}

Two asymptotic properties, each is associated with a scaling function, have been proposed in order to introduce generalized entropies to a large statistical system \cite{Thurner1}. Each scaling function is characterized by one critical exponent. These are $c$ for first and $d$ for second property, by which an equivalence class of entropies can be defined, uniquely, 
\bea
S_{c,d}[p] &=& \sum_{i}^{\Omega} {\cal A} \Gamma(d+1, 1 - c \ln p_i) + {\cal B} p_i, \label{eq:NewExtns1}
\eea
where $\Omega$ is the number of states, $\Gamma (a, b)=\int_{b}^{\infty}\, dt\, t^{a−1} \exp (−t)$ being incomplete $\Gamma$-function and ${\cal A}$ and ${\cal B}$ are arbitrary parameters. In its $\Omega \rightarrow \infty$ limit, each admissible system approaches one of these equivalence classes. It was concluded \cite{Thurner1} that the universality class $(c, d)$ not only characterizes e generic entropy, completely, but also specifies the distribution functions
\begin{eqnarray}
\varepsilon_{c,d,r}(x) = \exp\left\{-\dfrac{d}{1-c}\left[\mathtt{W}_k\left(B\left(1-x/r\right)^{1/d}\right) - \mathtt{W}_k(B)\right]\right\}, \label{eq:ps1}
\end{eqnarray}
where $\mathtt{W}_k$ is $k$-th branch of Lambert-{$\mathtt{W}$} function, which has real solutions at $k=0$ for all classes with $d\geq 0$ and at $k=−1$ for $d<0$. $B\equiv (1-c)r/[1-(1-c)r]\, \exp \left\{(1-c)r/[1-(1-c)r]\right\}$ with $r=(1-c+c d)^{-1}$. At $k=0$, the asymptotic expansion of Lambert-{$\mathtt{W}$} function reads
\bea
{\mathtt{W}}_{k=0}(x) &=& \sum_{n=1}^{\infty} \frac{(-1)^{n-1}\, n^{n-2}}{(n-1)!}\, x^n.
\eea

The properties of this new (non)extensivity entropy, Eq. (\ref{eq:NewExtns1}), lead to
\begin{eqnarray}
\dfrac{1}{1-c} &=& \lim_{N \rightarrow \infty} N\, \frac{\Omega^{\prime}}{\Omega}, \label{5} \\
d &=& \lim_{N \rightarrow \infty} \log \Omega\, \left(\dfrac{1}{N} \frac{\Omega}{\Omega^{\prime}}+c-1\right), \label{6}
\end{eqnarray}
while the number of microstates ($\Omega$) is related to the distribution function 
\begin{eqnarray}
\Omega(N) &=& \dfrac{1}{\varepsilon_{c,d}(-\varphi\, c\, N)}
  \exp \left\{ \dfrac{d}{1-c} \mathtt{W}_k\left(\dfrac{(1-c)\exp [(1-c)/c\, d]}{c\, d} \left[ \dfrac{\varphi\, c\, N}{r} \right]^{1/d}\right) \right\}, \label{eq:states1}
\end{eqnarray}
where $\varphi$ is given by 
\begin{eqnarray}
\varphi = \dfrac{d}{d N}\, S_g &=& \Omega^{\prime}\, \left(g(1/\Omega)-\dfrac{1}{\Omega}\, g^{\prime} (1/\Omega)\right). \label{scd}
\end{eqnarray}

Before introducing the grand-canonical partition-function, in which this generic (non)extensive statistics in incorporated, we want to implement this novel statistics in two different models, which not only reveal essential statistical properties but also have been utilized in characterizing various aspects of the high-energy particle-production.
 
\subsubsection{Ising model}
\label{sec:ising}

Ising model is a very simple spin model. It has been extensively utilized in high-energy collisions - among others - in determining the scaling properties of the hadron production \cite{ising1,ising2}. Accordingly, one might want to use its properties, such as, power-law scaling and intermittancy, and directly assumes that the given characteristics of the entropy can be directly applied on high-energy collisions and especially on lattice QCD thermodynamics. 

On a growing undirected random network of $N$ nodes connected by $L$ links (spin lattice), the interaction of two parallel (anti-parallel) spins contributes with $J^+\;$ $(J^-)$ to the total system energy. This  leads to \cite{Thurner1}
\begin{eqnarray}
E &=& \dfrac{L}{N(N - 1)}\, \left\{[n^+(n^+ - 1)+n^-(n^- - 1)] J^+ + 2 n^+ n^- J^-\right\} + \mu\, L = 2 \phi n^+(N - n^+)\, \Delta J.  \label{8}
\end{eqnarray}
where $\Delta J=J^- -J^+$ and $\epsilon=E/N$ being energy density per node. $\mu=-J^+$ is the energy cost for a link and $n^{+(-)}$ denotes the number of up (down) spins. The connectivity and connectancy of the network is defined as $k=L/N$ and $\phi =L/N(N-1)$, respectively. 
\begin{eqnarray}
n^+ &=& \dfrac{N}{2} \left( 1-\sqrt{1-\dfrac{2\epsilon}{k\Delta J}} \right) \simeq \dfrac{\epsilon}{2 \phi \Delta J}. \label{9}
\end{eqnarray}

The number of accessible states ($\Omega$) is given by the number of ways in which up-spins ($n^+$), for instance, are distributed over $N$ nodes, i.e., $\Omega=\tiny{(\begin{array}{c} N \\ n^+\end{array})}$. 
\begin{itemize}
\item  For constant connectivity $k$, then $\Omega=\tiny{(\begin{array}{c} N \\ a N\end{array})} \simeq b^N$, where $b=a^{-a}(1-a)^{a-1}>1$ and $0<a<1$. From Eqs. (\ref{5}) and (\ref{6}), $c=1$ and $d=1$, respectively. This leads to extensive entropy.
\item For constant connectivity $\phi$, $k=\phi\, N$ and $n^{+} \sim \epsilon /2 \phi \Delta J=const$. This follows that $\Omega \simeq  (N/n^{+})^{n^{+}} \exp(-n^{+}) \propto N^{n^{+}}$. From Eqs. (\ref{5}) and (\ref{6}), we find that $c=1-1/n^{+}$ and $d=0$. This means Tsallis $q$-entropy. 
\item If connectivity $k \propto N^\gamma$ with $0<\gamma<1$, then generalized entropies with $c=1$ and $d=1/\gamma$ likely have a super-exponential growth.
\end{itemize}
The values of $c$ and $d$ define essential statistical properties of this spin system. Accordingly, both connectivity and connectancy can be determined. Also, both extensivity and nonextensivity can be obtained from the proposed statistics. 

In the section that follows, we introduce another widely-used model based on counting, correlating and coalescence of micro hadron states, the quark-hadron coalescence model. Since the early 1980s, the coalescence approach found implications in nuclear collisions, from which recombination of deconfined quarks can be studied in the ALgebraic COalescence Rehadronization (ALCOR) model.

\subsubsection{Quark-hadron coalescence model}
\label{sec:coalescent}

Coalescence or recombination of constituent particles describes a very general process in a large number of systems ranging from femto- to astro-physics.  In particle physics,  the number of produced hadrons can be determined by the number of valence quarks (antiquarks) and gluons, which are mainly the constituents of those hadrons, multiplied by a coalescence coefficient ($C_{hadron}$) and a nonlinear normalization coefficient ($b_{q}$). In order to assure conservation of the quark numbers, the results should be normalized \cite{c3}.  Since the discovery of the baryon enhancement, this model successfully reproduced a huge amount of experimental data. The final hadron states are created in thermal equilibrium \cite{c44}. The spin degeneracy ($S_{hadron}$) and an additional normalization factor ($D^{hadron}=2\, S_{hadron}+1$) added to the model. 

The number of  meson and baryon states with meson-flavors ($i$ and $j$) and baryon-flavors ($i$, $j$ and $k$) is respectively given as \cite{c44,c9},
\begin{eqnarray}
N_M &=&D^M\; C_M(i,j)\; b_{q_{i}}\, N_{q_{i}}\; b_{\overline{q_{j}}}\, N_{\overline{q_{j}}},  \label{eq:mesonNumber}\\
N_B &=& D^B\; C_B(i,j,k)\, b_q(i)\, N_q(i)\; b_q(j)\, N_q(j)\; b_q(k)\, N_q(k). \label{eq:baryonNumber}
\end{eqnarray}
It is obvious that Eqs. (\ref{eq:mesonNumber}) and (\ref{eq:baryonNumber}) can be reformulated as sum over $i$,  $j$ and $k$ for each hadron with $n_f$ quark-flavors, so that the hadron multiplicity can be calculated and then compared with the experimental results \cite{c9,c99}.  

The conversion from nonlinear to linear rehadronization (or vice versa) is a straightforward exercise. Accordingly, the number of resulting hadrons is directly proportional to the product of the constituent quarks (with $n_f$ flavors) \cite{c5}. For instance, 
\bea
N_p  \propto N_q^{3}, \qquad
N_{\Lambda|\Sigma}  \propto N_q^{2}\, N_s, \qquad
N_\Xi  \propto N_q\,  N_s^{2}, \qquad
N_{\Omega}  \propto N_s^{3}. \label{eq:linear1}
\eea
It is obvious that both proportionality factor and the exponents of quark flavors are able to estimate the universality class  $d$, which is conjectured to be deduced from secondary scaling relation. The latter was proven for generic generalized entropies satisfying the first three Shannon-Khinchin axioms \cite{ShannonKhinchina,ShannonKhinchinb}, section \ref{sec:nonexp}, i.e., nonextensivity.

For the sake of simplicity, we assume that the system in composed of degenerate quarks and the hadronization procedure is very simple. The latter means constant connectivity. This minimizes the possibilities of forming hadrons from a concrete number of quarks. From Eqs. (\ref{eq:mesonNumber}) and (\ref{eq:baryonNumber}),  the quark numbers forming mesons and baryons is assumed to determine the number of the produced hadrons,
\bea
N_M \propto \frac{1}{2}\, N_q,  & \qquad &
N_B \propto \frac{1}{3}\, N_q. \label{eqLNMB}
\eea
Expression (\ref{eqLNMB}) gives the accessible hadron-states out of  $N_q$, $\Omega=\tiny{(\begin{array}{c} N_q \\ N_{M|B}\end{array})}$, which apparently can be reexpressed as functions of either $N_q$ or $N_{M|B}$ leading to $c=d=1$, i.e., BG extensive entropy. If the hadronization process (coalescence) takes place with
\begin{itemize}
\item a constant connectivity, then Eq. (\ref{eqLNMB}) leads to $d\, N_{N|B}\propto N_q$, with $\alpha={\cal O}(2)$ for bosons and $\alpha={\cal O}(3)$ for baryons. Then, the number of accessible states $\Omega\simeq b^{N_q}$ and the system is described by an extensive entropy, or

\item a constant connectancy, this  leads to $k\propto N_q$ and accordingly the number of accessible hadron states  $\Omega\simeq N_q^{N_{M|B}}$. The system is described by a nonextensive $q$-entropy, or 

\item a proportional connectivity (departing from the assumption of simplest confining procedure and applying relations as given in (\ref{eq:linear1})), then $k\propto N_q^{\alpha}$, leads to a super-equiexponential growing entropy with $c=1$ and $d=1/\alpha$. 
\end{itemize}
It is apparent that the last case sets limits on $\gamma$ as was predicted from Ising model.

\subsection{Generic (non)extensive partition function}
\label{sec:pfunction}

With the distribution function, Eq. (\ref{eq:ps1}), a partition function at vanishing chemical potential can be suggested. Assuming  classical gas of hadrons and resonances, then
\bea
\ln\, Z_{cl}(T) &=& V\, \sum_i^{N_{M|B}}\, g_i \int_0^{\infty} \frac{d^3\, {\bf p}}{(2\, \pi)^3}\; \varepsilon_{c,d,r}(x_i), \label{eq:PFcdr1}
\eea
where $V$ is the fireball volume and  $x_i=[({\bf p}^2+m_i^2)^{1/2}]/T$ characterizes $i$-th hadron or resonance dispersion relation. The conversion to quantum statistics is straightforward,
\bea
\ln\, Z(T) &=& \pm V\, \sum_i^{N_{M|B}}\, g_i \int_0^{\infty} \frac{d^3\, {\bf p}}{(2\, \pi)^3}\; \ln\left[1\pm\varepsilon_{c,d,r}(x_i)\right], \label{eq:PFcdr2}
\eea
where $\pm$ represent fermions and bosons, respectively. 

Recently, it was shown that the $q$-statistics seems to violate CPT symmetry that interprets holes among negative energy-states as anti-particle with the corresponding positive energy  \cite{tamas}. In order to fulfil the Kubo-Martin-Schwinger (KMS) relation \cite{kmsA,kmsB}, new quantum statistical distributions have been proposed. By using a deformed exponential function satisfying $\exp_{\kappa}(x) \cdot \exp_{\kappa}(-x)=1$, 
\bea
\exp_{\kappa}(x) &=& \left\{\left[1+ (b\, k\, x)^2\right]^{1/2} + b\, k\, x\right\}^{1/k},  \label{eq:nqsd}
\eea
where 
\bea
b=\frac{\left[\exp_q(x)\right]^k - \left[\exp_q(-x)\right]^k}{2\, k\, x},
\eea
$k=1/(1-q)$ and $q$-exponential function $\exp_q(-x)$ is given in Eq. (\ref{eq:exp}). The numerical estimation for lattice thermodynamics by using this $\kappa$-statistics (not shown here) still uses the so called Tsallis logarithmic function, i.e., conventional Tsallis algebra. Furthermore, finite chemical potentials should be taken into account, as well. We conclude that although fulfilling both KMS relation and CPT symmetry, new quantum statistical distributions, Eq. (\ref{eq:nqsd}), negligibly improves the productivity of lattice thermodynamics, for instance, by means Tsallis statistics.

For the sake of completeness, we emphasise that the low temperature deduced from the Tsallis fit for the transverse momentum distributions seems to depend on the used {\it ''specific''} formula. The {\it "standard"} formula, $Ed^3N/dp^3$, leads to Tsallis temperature ranging between $128$ and $150~$MeV \cite{ReferrRef1}, while another formula for invariant distribution, $d^3N/dp^3$  leads to much lower temperatures \cite{ReferrRef2}.

Coming back again to the claim that $q$-statistics {\it perfectly} reproduces the transverse momentum distribution at high energies (have been discussed in earlier section), Bialas had explained that it is incomplete to claim that the transverse momentum distribution of different produced particles at high energy is to be reproduced (actually fitted) from statistical models employing Tsallis algebra. There are two reasons \cite{bialas2015}:
\begin{enumerate}
\item Firstly, this claim seems not explaining the contradiction between statistical models with Tsallis algebra at high energy and the perturbative QCD approaches. The latter likely dominates such an energy region, while the earlier does not. 
\item Secondly, this claim seems to ignore the decay of statistical clusters, which can be scaled as power laws very similar to the Tsallis power-laws. The earlier ones exist over a wide range of transverse momenta, and thus, the statistical clusters decay might alone explain the surprising agreement between the measured transverse momenta and the calculations from statistical models even without Tsallis algebra. In other words, the power-law scaling might be due to decay of statistical cluster rather than Tsallis-nonextensivity.
\end{enumerate}

\section{Results and discussion}
\label{sec:results}

\begin{figure}[!h]
\includegraphics[width=5.5cm,angle=-90]{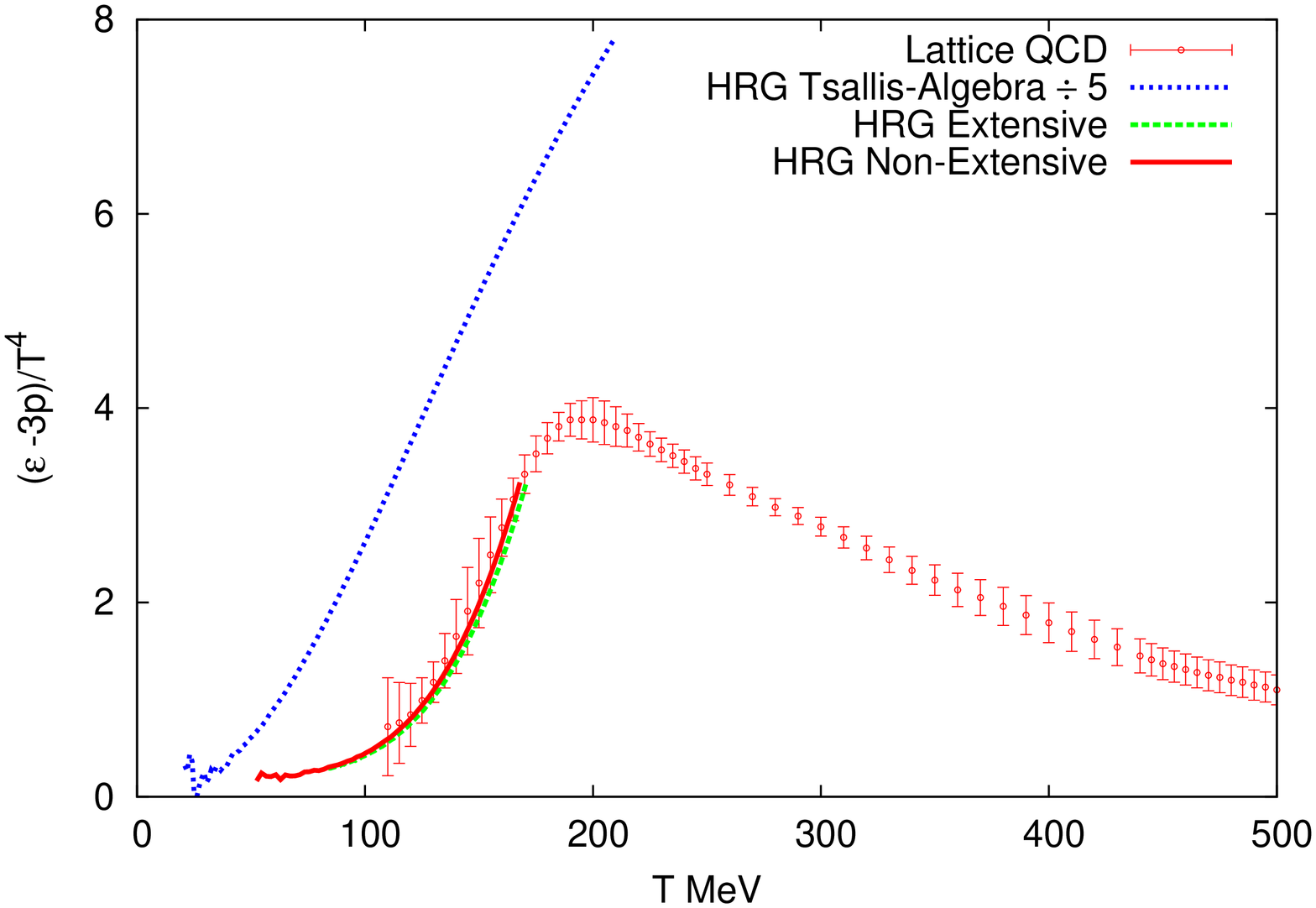}
\includegraphics[width=5.5cm,angle=-90]{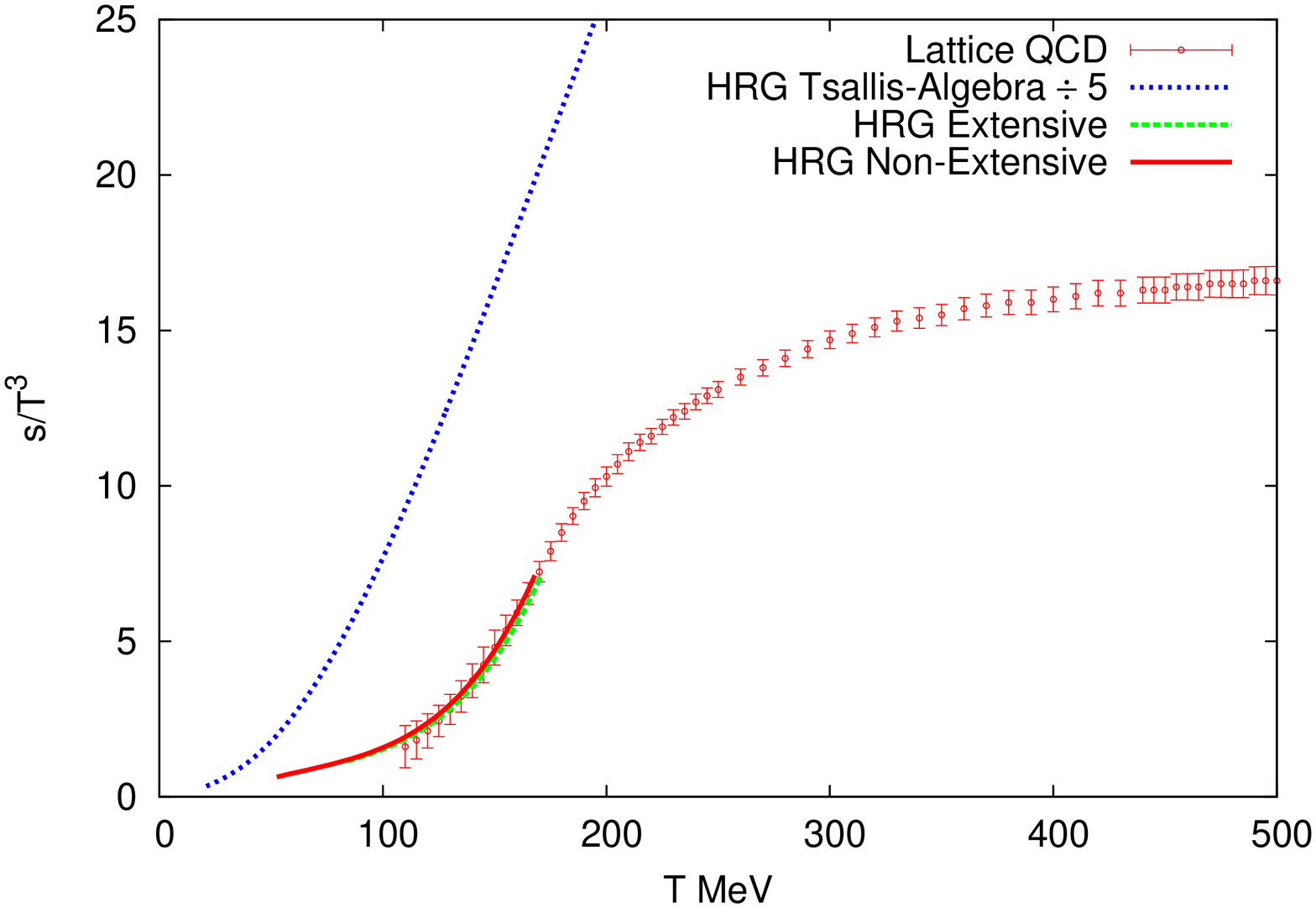}
\caption{A comparison between lattice QCD thermodynamics \cite{latticeQCD} (open circles with errorbars) and the corresponding calculations from HRG with BG- (dashed curve) and $q$-statistics (dotted curve). For a better illustration, the $q$-results are divided by $5$ in order to bring them to lower values. The solid curves represent HRG calculations in which nonextensivity is defined from the corresponding phase-space, section \ref{sec:pfunction}.  \label{fig:fg2a}}
\end{figure}

\begin{figure}[!h]
\includegraphics[width=8.25cm]{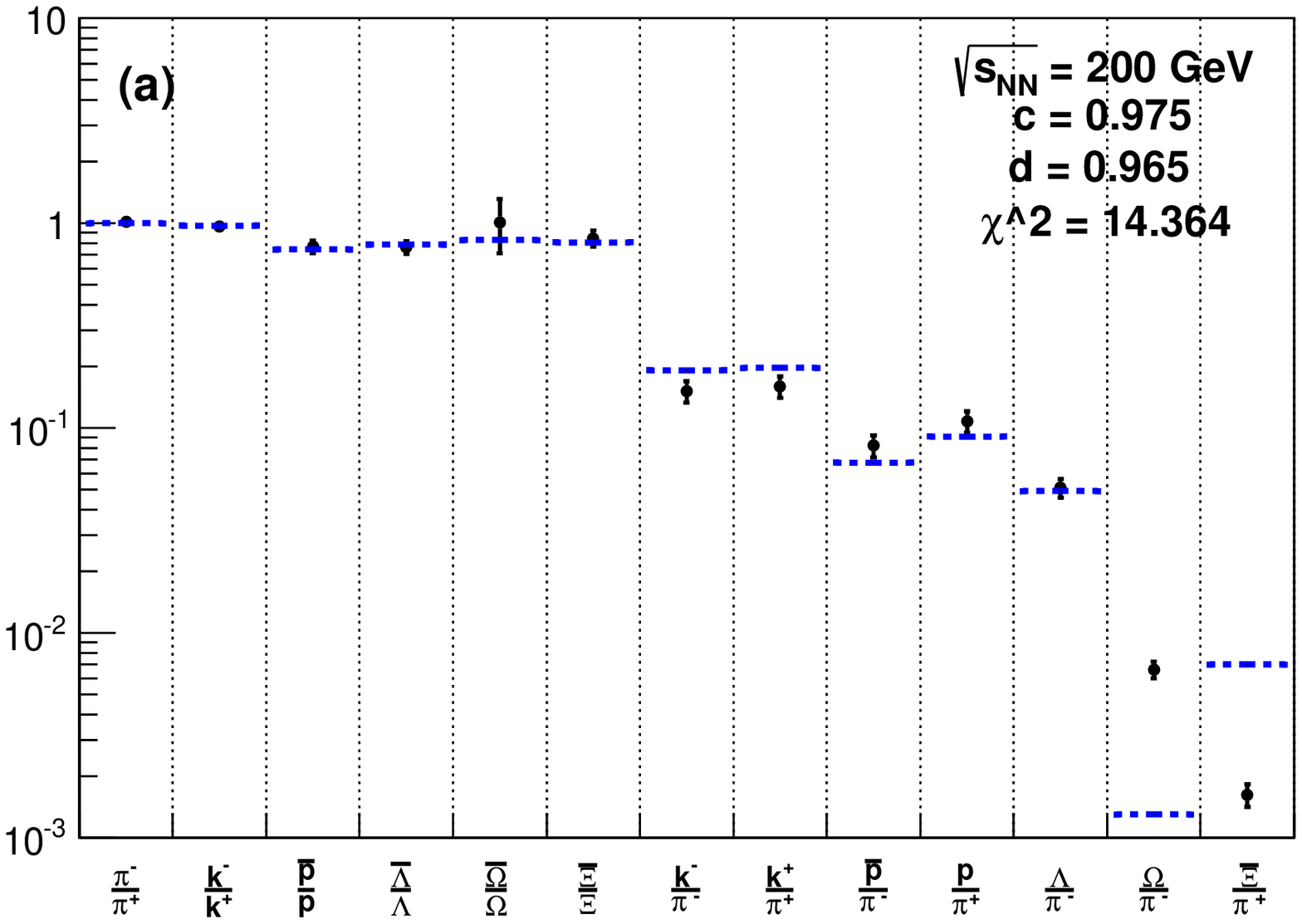}
\includegraphics[width=8.25cm]{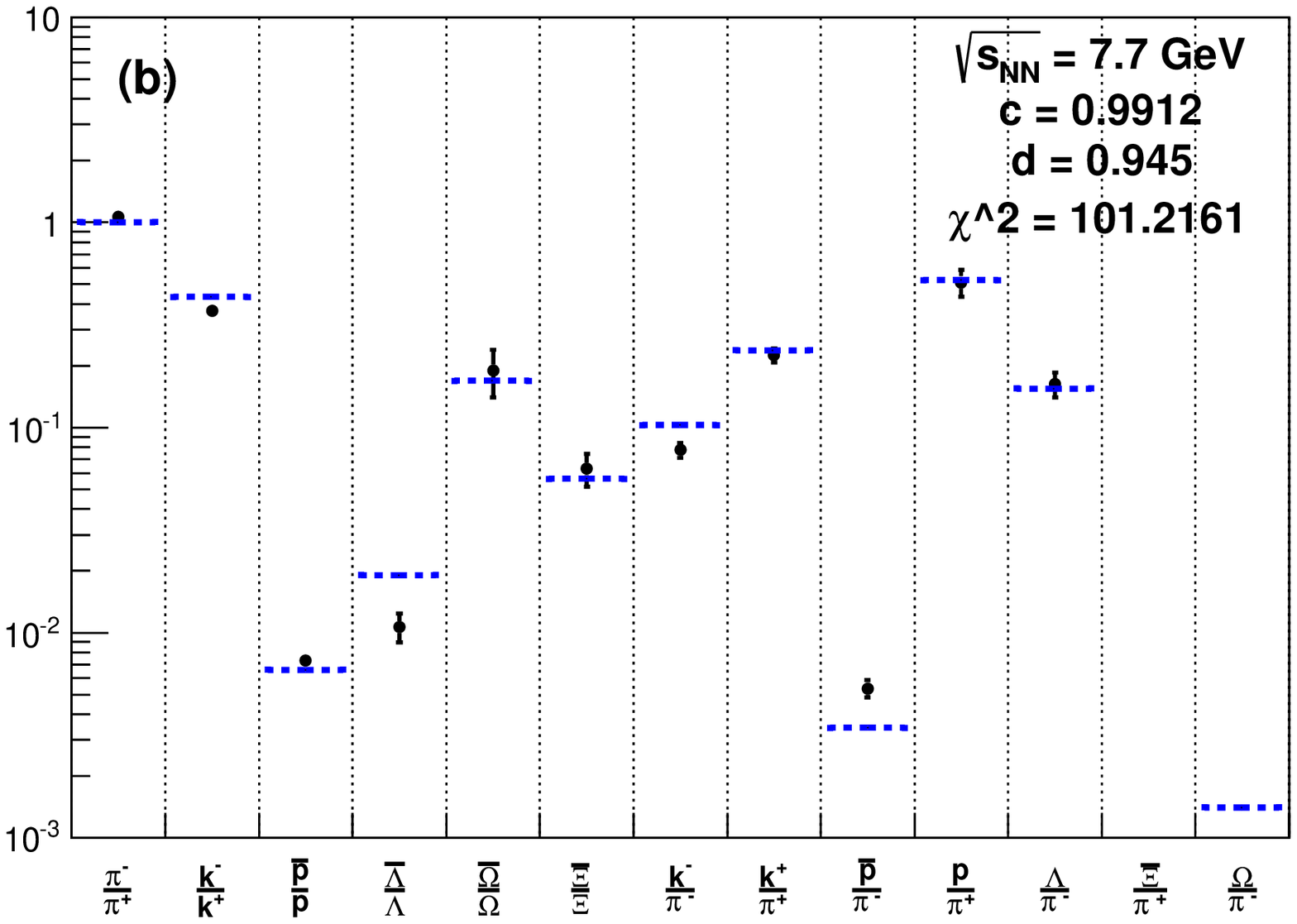}
\caption{A comparison between various experimentally measured particle ratios at $200$ [left-hand panel (a)] and $7.7~$GeV [left-hand panel (b)] given as solid circles and estimations from HRG, in which the generic (non)extensive statistics is implemented  (dashed line). The resulting $c$, $d$ and $\chi^2/dof$ are given in right top corers. \label{fig:fg2b}}
\end{figure}

Fig. \ref{fig:fg2a} depicts a comparison between lattice QCD thermodynamics \cite{latticeQCD} (open circles with errorbars); interaction measure ($(\epsilon-3 p)/T^4$) in left-hand panel and entropy density ($s/T^3$) in right-hand panel and the corresponding calculations from the HRG model with Boltzmann-Gibbs- (dashed curve) and $q$-statistics (dotted curve) at vanishing baryon chemical potential. It is obvious that HRG with Boltzmann-Gibbs statistics excellently reproduces the lattice results \cite{latticeQCD}, while HRG with $q$-statistics clearly overestimates both lattice thermodynamic quantities. For a better comparison, the $q$-results are divided by $5$. Furthermore, the qualitative dependence on temperature is not better than the quantitative comparison. The solid curves represent the HRG calculations in which generic (non)extensivity is defined from corresponding phase-space, section \ref{sec:pfunction}. The partition function in Eq. (\ref{eq:PFq1}) is replaced by Eq. (\ref{eq:PFcdr2}). Accordingly, 
\begin{itemize}
\item both lattice QCD thermodynamic quantities, interaction measure ($(\epsilon-3 p)/T^4$) and entropy density ($s/T^3$) are excellently reproduced, and
\item the resulting freeze-out temperatures agree well with the ones from the thermal (extensive) models at $q=1$, and
\item the resulting $c$ and $d$ parameters become very close to unity indicating extensivity, i.e., BG statistics.
\end{itemize}

Alternatively, nonextensive lattice QCD thermodynamics requires radical change in the lattice configurations. For instance, to obtain information about real-time quantities such as transport coefficients, analytical continuation of the imaginary-time calculations is essential, especially in anisotropic lattices with a smaller spacing in time than in space. Furthermore, designing lattice algorithms using random $\beta=1/T$ with sizeable variances still suffers from renormalization problems.  On the other hand, the effects of melting resonances, which increases with emerging into a continuum, could be a suitable ingredient to the HRG picture on the QGP side. In such thermal models, the nonextensive statistics might reproduce QCD thermodynamics in the anisotropic lattices.

Examples on different particle ratios measured in RHIC BES at $200~$GeV [left-hand panel (a)] and at $7.7~$GeV [right-hand panel (b)] are depicted in Fig. \ref{fig:fg2b}. These experimental results are compared with the statistical-thermal fittings, in which the proposed generic (non)extensive statistics (dashed lines) is incorporated. We conclude that the resulting nonextensivity parameters vary with the energy. At $200~$GeV, $c=0.975$ and $d=0.965$, while at $7.7~$GeV, $c=0.9912$ and $d=0.945$. These values refer to stretched exponentials \cite{Thurner1}. Also, the quality of both statistical fits gets worse with decreasing energy. At $200~$GeV, $\chi^2/dof=1.105$, while at $7.7~$GeV, $\chi^2/dof=7.785$. It is worthwhile to notice that the discrepancy with the experimental results is mainly in the strange particles species. This might be explained due to the fugacity factors, the quark occupation parameters and the excluded-volume corrections. The quark occupation parameters are assumed in equilibrium, i.e., both $\gamma_q$ and $\gamma_s$ are unity. No excluded-volume corrections are taken into consideration, i.e., HRG constituents are pointlike. The resulting freeze-out parameters are:
\begin{itemize}
\item at $200~$GeV, $T_{\mathrm{ch}}=148.05~$MeV and $\mu_{\mathrm{b}}=23.94~$MeV and
\item at $7.7~$GeV, $T_{\mathrm{ch}}=145.32~$MeV and $\mu_{\mathrm{b}}=384.3~$MeV.
\end{itemize}
We conclude that both sets of freeze-out parameters are obviously very compatible with the ones deduced from BG statistics \cite{Tawfik:2013bza}.

\section{Conclusions}
\label{sec:conclusions}

We have shortly reviewed the nonextensive statistics in high-energy collisions and highlighted that the resulting freeze-out temperatures from Tsallis-type nonextensive thermal-models are much lower than whose from both statistical thermal and lattice QCD approaches. We have discussed the attempts to extrapolate the Tsallis temperature to the well-known BG temperature. We assume that this difference is because of the improper implementation of Tsallis algebra on high-energy phenomena. With {\it ''implementation of Tsallis algebra''}, we mean one-to-one substitution of $\exp$ and $\ln$ functions by their Tsallis counter expressions. With {\it ''improper''} we mean that the replacement of extensive by nonextensive statistics should take into consideration  - among others - the possible modifications in the phase space. We suggest generic (non)extensive statistics, in which both Tsallis (nonextensive) and Boltzmann-Gibbs (extensive) are taken into consideration. These well-known types of statistics  describe very special systems. The degree of extensivity and nonextensivity can be dynamically determined from the phase-space distribution-function. We have applied this generic type of statistics to lattice QCD thermodynamics and particle ratios measured at various beam energies of RHIC BES. We found that the  lattice QCD thermodynamics is  extensive and likely additive (lattice techniques assume equilibrium QCD matter), while  the particle ratios are apparently nonextensive but not of Tsallis-type [the resulting $(c,d)\neq(q,0)]$.  We conclude that the resulting freeze-out parameters, temperature and baryon chemical potential, are compatible (not necessarily equal) with the ones deduced from BG statistics. 

Furthermore, we conclude that the HRG simulations, in which nonextensivity is defined from generic (non)extensive statistics determined from the distribution functions, reproduce well the lattice QCD calculations. Here, two asymptotic properties, each is associated with a scaling function, are utilized in order to classify a generalized entropy for such a high-energy system with large ensemble (produced particles) and strong correlations (various interactions). The resulting scaling exponents are $c\simeq 1$ and $d\simeq 1$. This means that the effective statistics is of Boltzmann-Gibbs type.  Also, HRG, in which extensivity is explicitly implemented, i.e., ideal gas of resonance constituents described by Eq. (\ref{eq:PFq1}), reproduces well the lattice QCD thermodynamics, as well. The resulting {\it critical} temperature is found compatible with the one determined from the extensive HRG model. In light of this, we conclude that the QCD simulations are extensive processes.

The well-reproduction of various particle ratios measured at $200$ and $7.7~$GeV by means of thermal models in which the generic (non)extensive statistics in utilized, shows that the resulting $c$ and $d$ refer to neither Boltzmann-Gibbs- nor Tsalis-type statistics. At $200~$GeV, $c=0.975$ and $d=0.965$, while at $7.7~$GeV, $c=0.9912$ and $d=0.945$. These values refer to stretched exponentials and asymptotically stable classes of entropy.


\end{document}